# Quantum Authentication Protocols for GSM

M. Hossientabar[1, 2], and B. Lari[3,*] H. Hassanabadi[4]

[1] Department of Physics, Ahvaz Branch, Islamic Azad University, Ahvaz, Iran
[2] Department of Physics, Khuzestan Science and Research Branch, Islamic Azad University, Ahvaz, Iran
[3] Department of Physics, Ahvaz Branch, Islamic Azad University, Ahvaz, Iran
[4] Physics Department, Shahrood University, Shahrood, Iran, P. O. Box 3619995161-316

E-mail: [1] m_ht51@yahoo.com, [3] behzadlari1979@yahoo.com[*] , [4] hha1349@gmail.com,

## Abstract

Security deficiencies and bugs in Authentication of SIM cards in Global Systems for Mobile (GSM) have led us to present new protocols for these networks using the principles of quantum cryptography. In this paper first, we provide a protocol for detecting and removing SIM card that has a copy, using three entangle particles source and quantum channel when the original SIM card and its copy simultaneously logging in the mobile network. Then, another protocol based on the use of quantum memory (which is embedded in SIM card) is presented. Both of these protocols can use to authenticate and remove SIM card that has a copy.

## 1. Introduction

Authentication as one of the most important issues of interest in mobile networks means identification of authorized subscribers. The close relationship between cryptography and authentication identifies the use of more secure methods [1-3]. At present, the use of longer code consists of a large string of numbers, more keys [4-7] and recently the use of n-field number theory [8] are the traditional methods of creating higher security in cryptography. All of these approaches sounded well until Peter Shore [9-10], by writing an algorithm, showed that even long stream codes could be broken to the prime numbers in a fraction of the time. Today, and after proposing this algorithm, we know that classical encryption methods with the advent of the next generation of quantum computers are not suitable methods for protecting information security [11]. Therefore, researchers have been using the quantum information theory and entanglement phenomenon to design cryptography protocols known as quantum cryptography that seems to have a much higher degree of security than classical ones. It seems that the use of the foundations of quantum science and a property called no cloning [12], can be effective in writing codes, with high security. In this regard, protocols such as BB84 [13-14], E91 [15], can be noted that highly secure protocols are used to supply quantum keys. Also in some papers, the three-qubit maximally entangled state GHZ, has been used for key distribution [5-6]. In this regard, we have also been discussing the use of the quantum principles to design protocols for Authentication on GSM [16-17]. Earlier, the BB84 protocol has been used in quantum authentication [18-19]. Now, with the benefit of the E91 protocol and the use of the source of three particles entangle state and also, the use of quantum memory, we want to authenticate the SIM cards. We know that many methods of identification are used in use of bank cards, electronic cards, wireless networks and mobile networks, all of which are in risk of disclosing security code and is subject to cybercrime attacks. In this paper, we have presented protocols that use quantum cryptography principles and attempt to prevent the use of copyrighted SIM cards (which are sometimes provided without the knowledge of the network and the main subscriber) on mobile networks. In the existing Global Systems of Mobile (GSM), sometimes for the purpose of eavesdropping and abuse, the second SIM is copied from the SIM card. This means that the SIM card can connect to the network without interrupting the AUC (simultaneously with the original SIM card) and eavesdrop on its conversation. This is especially true when both SIM cards are located at a network node or in the one antenna area. This kind of eavesdropping is different from an attack on information or a common eavesdropping on information networks that can be repaired by encrypting information in different ways. In this case, it is usually necessary to equip the SIM and the network to identify the subscribers, in such a way that it is confronted with a fake and copied SIM card. For the reasons mentioned above, we have been using quantum cryptographic methods to provide protocols with which the AUC can identify the copied SIM cards and prevent them from entering. In this regard, this paper is organized as follows. In Section 2, the Authentication Protocol is reviewed in commonly used classical GSM. In Section 3, we describe the E91 protocol and in Section 4, we use its principles and provide a protocol in which the source of three particles entangled state is used to identify and remove SIM cards that are duplicated. Section 5 provides another protocol in which SIM cards containing quantum memories are used, which, if copied, are detectable by the network. In Section 6, we will summarize the results of the previous sections. Before starting the sections, two points should be taken into consideration. First, in order to benefit from the protocols presented here, we need to design and create new tools such as quantum measuring devices, the source of three entangle particles and the embedding of quantum memories [20-21] in SIM. Second, the authentication has been classified into message authentication, identification, and key authentication. In the protocols presented here, attention is paid to the identification of SIM cards.

## 2. Authentication on Classical GSM

In this section, in order to better understand the Authentication on GSM we review the steps of this protocol. This can help us establish protocols based on the principles of quantum cryptography to identify the identity of the SIM cards in the future of mobile networks.

1. The subscriber (or SIM card) sends a request with International Mobile Subscriber Identity (IMSI).
2. For each IMSI, the network stores a $K_i$ number (which is exactly the same number stored on the SIM card). By applying this number and another randomly generated number called RAND in $A_3$ algorithm (which is specific to the network), generates another number called RES.
3. The network sends RAND to the SIM card and saves the RES number for comparison at a later stage of identification.

4. The subscriber uses $K_i$ and RAND sent by the network (following the request), and using these two numbers in the $A_3$ algorithm, generates a number that we call it XRES.
5. In the final step, the SIM card sends the XRES number to the network, and the networks compare it with the RES. If the two numbers are equal, the SIM card's identity is confirmed otherwise, the SIM card will not allow the network to be used.

At the end of this section, it is necessary to state that the $A_3$ algorithm is the authentication algorithm in the GSM security model. The $A_3$ algorithm can be any kind of encryption because of the design of the GSM system allows an operator to choose any algorithm. The $A_3$ algorithm gets the RAND from the Mobile Switching Centers (MSC) and the secret key $K_i$ from the SIM as input and generates a 32-bit output, which is the RES response. Both the RAND and the $K_i$ secret are 128-bits long. The $A_3$ algorithm can be typed as a one-way hash function. Generally, one-way hash functions produce a fixed-length output given an arbitrary input. Secure one-way hash functions are designed such that it is computationally unfeasible to determine the input given the hash value, or to determine two unique inputs that hash to the same value. Since the quantum protocols are presented in the next section, the results of the measurement are recorded with classical bits, it is possible to use the $A_3$ algorithms for $QK_j$, similar to those used today in telecommunications.

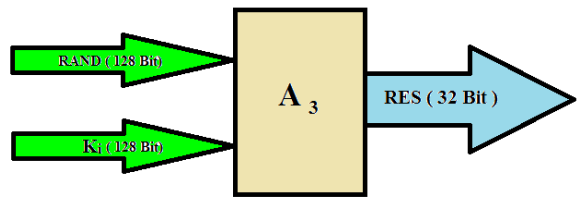

## 3. Protocol E91

In this section, we describe the E91 protocol that is based on the two entangled particles shared from a source to Alice and Bob to distribute a secret key. Here, with an overview of this method, the protocols are based on the principles of quantum cryptography and no-cloning property in quantum mechanics, to identify the SIM cards in the next generation of mobile networks. We call these protocols, quantum GSM (QGSM).

1. An entangle particles source produces an entangled state such as Bell states and shares it between Alice and Bob.
2. Alice and Bob each one does measurements on their part in preferred bases such as X or Z separately.
3. After measurements, Alice and Bob, using a classical channel, merely inform their measurement bases without announcing the measurement result.

4. If their measurement bases are the same at each stage, they will save the result of their measurement in classical memory using classical bits otherwise they will eliminate the results.
5. To avoid eavesdropping, they announce some of their measurement results.
6. Alice and Bob can use these stored bits as the quantum key to encrypt shared information among themselves and through the classical channel.

By using the protocols in the previous sections, in the sequence, protocols can be provided for the next generation of mobile networks.

## 4. SIM card identification protocol using three entangle particle

In the protocol outlined in this section, a source of three entangled particles will be used. It is also assumed that the network in addition to identifying the SIM card by the method applied in GSM means using $K_i$ in the $A_3$ algorithm on any SIM card and network, and comparing RES$_i$ and XRES$_i$ results. It also uses an exclusive quantum key (Which is called $QK_j$) that is shared between each SIM card and network and pair of $(K_i, QK_j)$ separately are applied in algorithm $A_3$ and produce a pair of $(RES_i, RES_j)$ in the network. Also, a pair of $(XRES_i, QXRES_j)$ is generated on each SIM card. By comparing these results, identification is done. $j \in \{1,2\}$ is the index of the SIM cards.

1. Each SIM card (original or copied) sends network identification requests through a classical channel with IMSI.
2. The source of entanglement sends three entangled particle to the SIM cards and Authentication Center (AUC), which is embedded in the network.
3. Each SIM card and AUC separately perform measurements in the desired bases $X$ or $Z$ on their particles sent from the entanglement source and record their result with a classical bit.
4. Each SIM card can share their measurement bases with the AUC through a classical separate channel. If the bases are the same, the result of the measurement (which is registered with a classical bit) is stored and used as a quantum key otherwise the results will be discarded.
5. If the network finds that the measurement basis are the same in all three parts, and in the X direction, by sending a Null message through the classical channel to each of the parts, which we will explain in the sequence. However, if the measurement direction for all three parts is Z, the result will be registered.
6. The AUC and each SIM card use both $(K_i, QK_j)$ and apply them separately in $A_3$ and generate pairs $(RES_i, RES_j)$ and $(XRES_i, QXRES_j)$ respectively. By comparing these results, the network can authenticate the SIM cards.
7. The network will only notice the existence of a copy of the SIM card due to the difference in results with the j-index and the identical results of the i-index for the dual SIM card.

### 4.1. Properties

(a) **Randomness**

As it is stated in this protocol, in the production and use of the quantum key, it is completely random and dynamically evident by the momentary production of entangled particles and their measurement. Therefore, the protocol is quite dynamic in this regard.

(b) **Unidirectional authentication**

The authentication can be categorized into unidirectional authentication and bidirectional authentication. In reviewing the above protocol and in accordance with its clauses, identification is performed by the AUC and is therefore unidirectional.

### 4.2. Null Massage

The Null is a message from AUC to SIM cards. This message is sent when the basis of measurement is X in all three parts. The AUC must operate in such a way that the following principles are not violated. (a) If any of the SIM cards are logged in to the network alone, the network shall not prevent them from entering. (b) To maintain a random measurement, the measurement basis X and its results shall not be removed for any reason, since measuring in a particular direction such as Z will cause the copied SIM card to be measured only in the same direction and always get the correct quantum key and this will prevent identification. As indicated, the Null message is sent when the AUC determines that the basis of the measurement is X. The result of sending this message will be such that, in the event of mismatch between the results recorded in different parts, they are identical and applied in the recording of the quantum key and by this, the principles stated above are respected. In later sections for GHZ mode and W mode, the result of sending this message and its operation will be explained.

### 4.3. Possible scenarios

Suppose a copy of the SIM card is provided. The question is how can the AUC know about the existence of a SIM card that has copied and remove both of them? In the sequence and in response to this fundamental question, it is assumed that the $A_3$ algorithm and $K_i$ key associated with the original SIM card are fully copied. The Possible scenarios in identifying and not identifying copied SIM cards are in following

1. If each SIM card is unintentionally requested with its IMSI, the AUC can't detect copying of each of them by applying the above steps, and the SIM cards can be logged in and communicated on the network.
2. When both SIMs at the same time request, the network encounters a situation for two requests, in which the results labeled with the index "i" are the same, but the results that are labeled with the "j" index are different. The reason is the quantum key can't be copied. So it can detect copied SIM cards. Even if in rare cases and randomly, the quantum key for the two SIMs is the same, there will be no problem, because in general and in other requests, this key is different and it is possible to determine if the SIM card is copied.
3. If the copied SIM card attempts to eavesdrop the measurement bases, since the quantum key is different from the original SIM card's quantum key, the AUC can still be informed of copying the original SIM card.
4. If the original SIM card using a classic channel, sends its quantum key to other SIM card, the dual sim card can enter in to the network without any problems.
5. A classical channel can be established between two SIM cards such that, if one SIM card is requested, another SIM card can be used to prevent it from being logged in to the network. It can be imagined that, through this classical channel, the quantum key generated by SIM1, which is stored in classical bits, is transferred to SIM2 such that it can be logged in at the same time. Such results in the inability of AUC to detect a SIM card that has a copy.

Items 4 and 5 are only possible with a wireless connection at short distances between the original and the copied SIM cards, which is not practically justifiable. Therefore, when these two SIM cards are not in the wireless connection range and simultaneously request to the network, the authentication a SIM card that has a copy, is possible by the AUC.

### 4.4. Distribute the quantum key for the protocol

In this section, first we discuss how to share the quantum key in each one of the channels and the defects that should be identified to prevent the production of the wrong key. To simplify, we call the original SIM card with SIM1, the copy of SIM card with SIM2, the authentication center of network with the AUC and the source of entangled particles (which can share three entangled particles) by SEP. Now imagine the source of entangled particles, share the GHZ-state (for example $|\psi\rangle = \frac{1}{\sqrt{2}}(|000\rangle + |111\rangle)$), between three parts. Here, the ket $|ijk\rangle$ is a state that has shared from the entangle source and i, j and k is indexing for SIM1, SIM2 and AUC respectively. We should note that there is no difference between the SIM1 and its copy. Now we introduce the following projective measurement:

$$\mathcal{M}_{p\pm} = |S_p; \pm\rangle\langle S_p; \pm| \quad \text{where } p \in \{x, y, z\} \quad (1)$$

Suppose that at first emission of particles from the source, measurements in the desired directions X, Z, and Z are arranged in sequence AUC, SIM1, and SIM2 and relative results of measurements are $|+\rangle$, $|0\rangle$ and $|0\rangle$ respectively. We need to point out, the timing sequence is important in measuring each part. If this sequence changes, then the final result of the measurement will vary. But what matters most is whether or not the bases are identical and if the bases are the same, the results of the measurement are stored with the classical bits to be used to make the key. Of course, the most important thing is that the network should be informed of the shared state between the three parts (here the GHZ-state). Because, as we will see later, if all three measurement bases are the same (this can be detected by the network) and in the case of working with GHZ-state or W-state, only if every three measurement bases are Z, the result can be stored (Of course in the case of w-state, since the AUC knows the state sent by the source, it must flip the results of its own measurement.). In the case that all three are measured at the X-bases, the network should send a Null message through the classical channel to each one of the SIM cards. The reason is that the saved results are not identical between each SIM card and network. The results of the measurements mentioned above can be written using the projective measurement.

$$|\psi'\rangle = (I_1 \otimes I_2 \otimes \mathcal{M}_{X+})|\psi\rangle = \frac{1}{\sqrt{6}}(|00\rangle_{12} + |11\rangle_{12}) \otimes |+\rangle_3$$

$$|\psi''\rangle = (\mathcal{M}_{Z+} \otimes I_2 \otimes I_3)|\psi'\rangle = \frac{1}{\sqrt{6}}|00\rangle_{12} \otimes |+\rangle_3 \qquad (2)$$

$$|\psi'''\rangle = (I_1 \otimes \mathcal{M}_{Z+} \otimes I_3)|\psi''\rangle = \frac{1}{\sqrt{6}}|00\rangle_{12} \otimes |+\rangle_3$$

Where $I_m$ is the matrix of $2 \times 2$ for $m \in \{1,2,3\}$, or equivalent $m \in \{SIM1, SIM2, AUC\}$, and the ket $|\pm\rangle_m$ is the Eigen state of $X_\pm$ for part $m$. In the rest of this section, we also show the communication channel (1,3) with $A$ and the communication channel (2,3) with $B$. Therefore, in the sequential measurements indicated above, the classic bits $(0,0,0)$ will eventually be registered for $(1,2,3)$, regardless of its coefficients. But since the bases of measurement for Channel $A$ and Channel $B$ are not the same, these results will not be used to make the quantum key in each of the channels mentioned, and the results will be deleted. Now consider the case that the measuring bases, the ordered triplets (Z, X, Z) and the timing sequence of measurements are similar to Eq. (2). Then the classical result $(0,0,0)$ will be registered by each of the parts according to the following calculations. In this case, only Channel $A$ will uses the results registered as a key due to the similarity of its bases with the measurements bases of the network.

$$|\psi'\rangle = (I_1 \otimes I_2 \otimes \mathcal{M}_{Z+})|\psi\rangle = \frac{1}{\sqrt{2}}|00\rangle_{12} \otimes |0\rangle_3$$

$$|\psi''\rangle = (I_1 \otimes \mathcal{M}_{Z+} \otimes I_3)|\psi'\rangle = \frac{1}{\sqrt{6}}|0\rangle_1 \otimes |0\rangle_2 \otimes |0\rangle_3 \qquad (3)$$

$$|\psi'''\rangle = (\mathcal{M}_{X+} \otimes I_2 \otimes I_3)|\psi''\rangle = \frac{1}{\sqrt{6}}|+\rangle_1 \otimes |0\rangle_2 \otimes |0\rangle_3$$

In Table-1, the possible measurements for the state $|\psi\rangle$ are summarized. In the red column, Channel $A$ and $B$ (because of difference between their measurement bases with AUC) cannot register any bit to make a quantum key. In the blue column, the bit "0" is registered in channel the $A$, and in the green column, the bit "0" is registered in channels $A$ and $B$. Just in channel $B$, the bit "1" is registered for the quantum key. The gray column is a Null-column and the results recorded by each SIM card must be changed according to the Null message, and then used to make the quantum key. In the following, we will show the Null column for the GHZ-state and for one of the W-states. The timing sequence in the measurement is again the same, which is, of course, quite arbitrary.

### (a) Null for GHZ-state

On a successive measurement, first, the AUC, then SIM1 and finally SIM2, perform their measurement in the X direction and obtain the results $|-\rangle, |-\rangle$ and $|+\rangle$ respectively. Which can be represented by projective measurement.

$$|\psi'\rangle = (I_1 \otimes I_2 \otimes \mathcal{M}_{X-})|\psi\rangle = \frac{1}{2}(|00\rangle_{12} - |11\rangle_{12}) \otimes |-\rangle_3$$

$$|\psi''\rangle = (\mathcal{M}_{X-} \otimes I_2 \otimes I_3)|\psi'\rangle = \frac{1}{\sqrt{6}}|-\rangle_1 \otimes |+\rangle_2 \otimes |-\rangle_3 \quad (4)$$

$$|\psi'''\rangle = (I_1 \otimes \mathcal{M}_{X+} \otimes I_3)|\psi''\rangle = \frac{1}{\sqrt{6}}|-\rangle_1 \otimes |+\rangle_2 \otimes |-\rangle_3$$

As it can be seen, when such a timekeeping sequence is observed in the measurement, the third measurement carried out by the SIM card will inevitably project to $|+\rangle$ and therefore, the number in this measurement cannot be recorded for making the quantum key between AUC and SIM2. Such a problem could have occurred in the case of registering the key between SIM1 and AUC depending on the sequence of measurements (the other cases in Table 2 that need to send a Null message are listed). Therefore, the network, by knowing that the entangled state which is shared is GHZ-state, can send a Null-message through the classical channel to both SIM cards as follows: (1) If the AUC is measured in X-basis and obtains the result of "+" (or "0"), it will then notify the SIM cards that if they also measured in X-basis and the result of "_" (or "1"), flip it, then use it in making the key. (2) If the AUC is measured in X-basis and obtained the result of "-" (or "1"), then it will notify the SIM cards that if they also measured in X-basis and the result of "+" (or "0"), flip it, then use it in making the key. It can be explained in accordance with Table 2. If AUC measurements and SIM cards are X basis then: (a) for the state $|+++\rangle$, no AUC message is sent to SIM cards, and the result (1, 1, 1) is recorded, (b) for $|--+\rangle, |+--\rangle$ and $|-+-\rangle$, after sending the Null message, in all three cases the result (1, 1, 1) is recorded. This means that the green columns for the X basis are still green in this table and the gray columns will turn into green columns and thus will not affect the measurement. As mentioned, green columns are used to make key, but have no effect on identification, and only yellow and blue columns are affected. It can also be seen that this Null-message has no contradiction with the two principles stated above.

### (b) Null for W-State

For $|\psi\rangle = \frac{1}{\sqrt{2}}(|010\rangle + |101\rangle)$, if the same as part (a), consider the same time sequence in the measurement for the X-direction. Then the recorded results with classical bits will again be (1.0.1) for (AUC, SIM1, SIM2). It can be illustrated with projective measurements. It can be seen that, after measuring the first and second measure, when the SIM card 2 finally measures, it obtains the state $|+\rangle$. In this case, the same instructions as in the case of the GHZ- state can be used to change the results of the measurement and then apply the quantum key.

$$|\psi'\rangle = (I_1 \otimes I_2 \otimes \mathcal{M}_{X-})|\psi\rangle = \frac{1}{\sqrt{2}}(|01\rangle_{12} - |10\rangle_{12}) \otimes |-\rangle_3$$

$$|\psi''\rangle = (\mathcal{M}_{X-} \otimes I_2 \otimes I_3)|\psi'\rangle = \frac{1}{\sqrt{6}}|-\rangle_1 \otimes |+\rangle_2 \otimes |-\rangle_3 \quad (5)$$

$$|\psi'''\rangle = (I_1 \otimes \mathcal{M}_{X+} \otimes I_3)|\psi''\rangle = \frac{1}{\sqrt{6}}|-\rangle_1 \otimes |+\rangle_2 \otimes |-\rangle_3$$

| $|\psi\rangle$ | Measurement Bases and Registered Classical Results | | | | |
|---|---|---|---|---|---|
| | Z | Z | Z | X | X |
| SIM1 | "1" | "0" | "0" | "1" | "1" |
| | X | Z | Z | Z | X |
| AUC | "1" | "0" | "0" | "1" | "1" |
| | Z | X | Z | Z | X |
| SIM2 | "1" | "0" | "0" | "1" | "0" |

Table 1. Measurement bases and registered results for different part in GHZ case.

### 4.5. An example of using the three-particle GHZ-State protocol

First, assume that each of the SIM cards separately and at different times, through the classical channel send a request. Suppose that $K_i$ is a number based on 2, such as 0110 and suppose, for the captured of the twenty state $|\psi_i\rangle$ that are sent from the source, $QK1 = 01011$ as a key in the channel A, and $QK2 = 10110101$ as a key for the channel B, are recorded. In the simplest case, assume that $A_3$ is an algorithm for summation and RAND = 10110. It can be seen that for both channels, regardless of synchronization or non-synchronization of requests, for the original SIM card and its copy, $RES_1 = RES_2$ but $QXRES_1 \neq QXRES_2$. Table 2, presents the measurements and their results in a completely random sequence of AUC, SIM1 and SIM2. Then, using the keys distributed above and following the protocol, which we will see in the schema below, we will keep informed of the copying of the SIM cards in progress and remove them to prevent possible eavesdropping.

| State | | $|\psi\rangle_1$ | $|\psi\rangle_2$ | $|\psi\rangle_3$ | $|\psi\rangle_4$ | $|\psi\rangle_5$ | $|\psi\rangle_6$ | $|\psi\rangle_7$ | $|\psi\rangle_8$ | $|\psi\rangle_9$ | $|\psi\rangle_{10}$ | $|\psi\rangle_{11}$ | $|\psi\rangle_{12}$ | $|\psi\rangle_{13}$ | $|\psi\rangle_{14}$ | $|\psi\rangle_{15}$ | $|\psi\rangle_{16}$ | $|\psi\rangle_{17}$ | $|\psi\rangle_{18}$ | $|\psi\rangle_{19}$ | $|\psi\rangle_{20}$ |
|---|---|---|---|---|---|---|---|---|---|---|---|---|---|---|---|---|---|---|---|---|---|
| SIM1 | MB | $S_x$ | $S_z$ | $S_x$ | $S_z$ | $S_z$ | $S_x$ | $S_z$ | $S_z$ | $S_x$ | $S_x$ | $S_z$ | $S_x$ | $S_z$ | $S_z$ | $S_x$ | $S_z$ | $S_x$ | $S_x$ | $S_z$ | $S_x$ |
| | Rs | $|-\rangle$ | $|1\rangle$ | $|+\rangle$ | $|0\rangle$ | $|0\rangle$ | $|-\rangle$ | $|1\rangle$ | $|0\rangle$ | $|1\rangle$ | $|-\rangle$ | $|1\rangle$ | $|+\rangle$ | $|0\rangle$ | $|+\rangle$ | $|-\rangle$ | $|1\rangle$ | $|+\rangle$ | $|-\rangle$ | $|0\rangle$ | $|-\rangle$ |
| | CB | 1 | 1 | 0 | 0 | 0 | 1 | 1 | 0 | 1 | 1 | 1 | 0 | 0 | 0 | 1 | 1 | 0 | 0 | 0 | 1 |
| AUC Unit | MB | $S_x$ | $S_x$ | $S_x$ | $S_x$ | $S_z$ | $S_x$ | $S_z$ | $S_z$ | $S_x$ | $S_z$ | $S_x$ | $S_z$ | $S_z$ | $S_z$ | $S_z$ | $S_z$ | $S_x$ | $S_z$ | $S_x$ | $S_x$ |
| | Rs | $|-\rangle$ | $|+\rangle$ | $|+\rangle$ | $|+\rangle$ | $|0\rangle$ | $|+\rangle$ | $|1\rangle$ | $|0\rangle$ | $|-\rangle$ | $|0\rangle$ | $|+\rangle$ | $|1\rangle$ | $|+\rangle$ | $|0\rangle$ | $|1\rangle$ | $|+\rangle$ | $|-\rangle$ | $|1\rangle$ | $|-\rangle$ | $|-\rangle$ |
| | CB | 1 | 0 | 0 | 0 | 0 | 0 | 1 | 0 | 1 | 0 | 0 | 1 | 0 | 0 | 1 | 0 | 1 | 1 | 1 | 1 |
| SIM2 | MB | $S_x$ | $S_z$ | $S_x$ | $S_x$ | $S_z$ | $S_x$ | $S_z$ | $S_x$ | $S_x$ | $S_x$ | $S_x$ | $S_z$ | $S_z$ | $S_x$ | $S_z$ | $S_x$ | $S_x$ | $S_x$ | $S_z$ | $S_z$ |
| | Rs | $|+\rangle$ | $|0\rangle$ | $|+\rangle$ | $|+\rangle$ | $|0\rangle$ | $|-\rangle$ | $|1\rangle$ | $|-\rangle$ | $|-\rangle$ | $|-\rangle$ | $|+\rangle$ | $|1\rangle$ | $|0\rangle$ | $|+\rangle$ | $|1\rangle$ | $|1\rangle$ | $|-\rangle$ | $|-\rangle$ | $|0\rangle$ | $|0\rangle$ |
| | CB | 0 | 0 | 0 | 0 | 0 | 1 | 1 | 1 | 1 | 1 | 0 | 1 | 0 | 0 | 1 | 1 | 1 | 1 | 0 | 0 |

**Table 2** - Quantum Key Distribution for the Three Entangle Particle (with GHZ-state) Authentication Protocol. Blue Column is Key Distribution on Channel $A$. Yellow Column is Key Distribution on Channel $B$. In Green Column, the same keys is Recorded on Channels $A$ and $B$. In the red column, because of the lack of uniformity of the measurement bases, and despite the uniformity of the classical bit recorded on both channels, a number is not stored for making the key. The gray column shows the Null column. In this table, MB, CB and RS stand for Measurement Bases, Classical Bit and Results, respectively.

In Figure 1, a schematic of authentication protocol using three entangle particles is shown. In this figure, $M_P$ is a Projective Measurement-Unit. In this protocol, it is necessary to design a quantum measuring instrument in each of the triples parts and create a quantum channel between them.

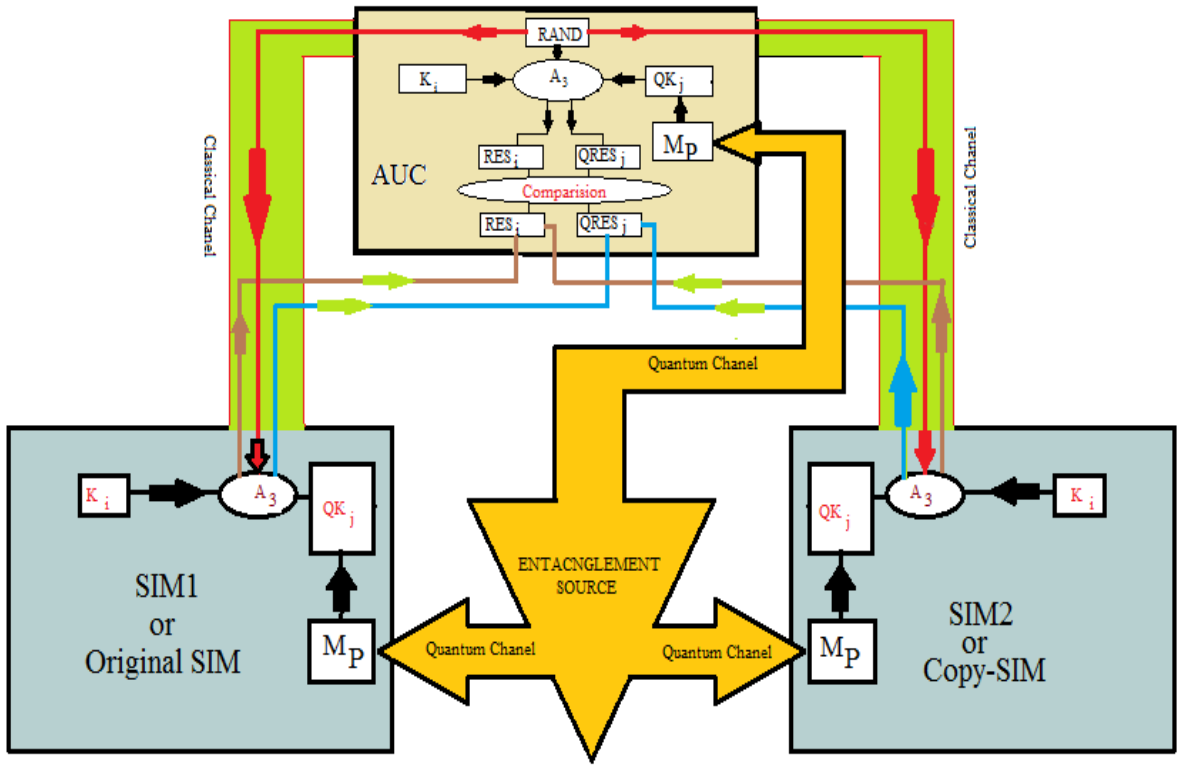

**Figure 1.** A schema of the protocol identifying and removing copied SIMs using the source of three entangle particle.

In the above protocol, which is basically based on quantum cryptography, we have used both classical and quantum channels in identifying the SIM card, like any other well-known protocols such as E91. But, as stated in Section 3, the E91 protocol is written for two particles, while we have written it for three particles. In addition, using the integration of protocols in the current telecommunication network to identify the SIM card, with the E91 protocol, we can use a new method to identify the identity of the SIM card to be copied, as described in this section. We also know that, the classical channel is used to compare the bases of measurements, so if the bases are the same, the results of the quantum measurement are recorded. Also, the results of quantum measurements, as stated, are stored with classical equivalent bits (see Table 2).

### 4.6. The CNOT Attack

The purpose of this section is to see whether the CNOT attack is affecting the process of identifying the SIM card copied using the identification protocols presented in this paper. First of all, the CNOT attack must be known. The eavesdropping model in this case is as follows, where the untrusted third-party Eve will try to obtain the information. Eve will take the qubits from Alice and Bob and put each one of them in the control input of a CNOT gate and she will supply $|0\rangle$ in the target. The outputs corresponding to the control qubits of the CNOT gates will be measured in the Bell basis by Eve and the result will be communicated to Alice and Bob. Eve stores the output corresponding to the target in her quantum memory.

Then Alice and Bob will go for public discussion to announce their bases. Knowing these, Eve will try to extract information from the outputs corresponding to the target qubits of the CNOT gates. Now imagine that the hacker will penetrate one of the channels A or B (Fig. 1) using the CNOT attack and get the quantum key of that channel. But paying attention to two points is important. First, the hacker gained the quantum key with probability, and not with certainty [22-23]. Second, the hacker cannot, with this method, simultaneously obtain the key of both channels (due to the distance between the two SIM cards). However, the protocol outlined in this section aims to detect and prevent copied SIM cards, which is obtained by mismatching the encryption keys on the two channels A and B. It may be argued that the key will be obtained on one of the channels and will be available to the other channel. As it is known, the spatial distance makes it impossible. Therefore, it seems that the hacker, in practice, cannot prevent the identification of such SIMs by using a CNOT attack.

## 5. SIM card authentication protocol using quantum memory

Quantum memory which is reported by many researchers, is a tool for storing information for an unlimited time, and then read out the information [21], [24-25].The construction of quantum memories based on the theory of quantum dots [26-29] is being investigated. Of course, in manufacturing memory, the amount of storage time is always noticeable, with no loss of data, with the least energy consumption and the speed of performance in storing and retrieving information. Here, we intend to provide a simple protocol for the use of the quantum memory which is embedded in the SIM cards to be able to identify the SIM cards by the network. In this way, the network can use this protocol to authenticate SIM cards that are likely to be used to eavesdrop the original SIM card and prevent them from entering the network. Now suppose the SIM cards have quantum memory and before assigning the SIM card to the subscribers at the Home Location Register (HLR), a source of entangled state share one of the Bell states as $|\psi\rangle = \frac{1}{\sqrt{2}}(|00\rangle + |11\rangle)$ between AUC and SIM cards. The particle that reaches the AUC is randomly measured in the X or Z bases, the bases and results of the measurement are stored in the AUC, and the particle that is sent to the SIM card is stored in the SIM card without measurement and within the quantum memory. Suppose a large number, for example, $N$ particles, is stored (which is sent by the source of entangled particles) that are not yet measured on them in the SIM card's quantum memory. With this assumption, we can propose the following protocol for quantum authentication in the mobile network using the quantum memory embedded in the SIM card.

1. The SIM card for entering the network sends a request with the IMSI number to the AUC.
2. The AUC, in accordance with the IMSI has been sent from the SIM cards, will ask them to measure the quantum states (spouse for the sites are labeled and stored on the SIM card with n to n + m) in one of the bases X or Z, and then send the results accompanied by the measurement bases for the AUC.

3. The AUC compares the bases used by the SIM card with its measurement bases that it has previously used. The result is deleted in sites where the bases are different. The result is stored in sites where the bases are the same.
4. In the recall of sites with similar labels, if the result of measuring the AUC is equal to the result of measuring the SIM card and the measurement bases, the SIM card will be authenticated and allowed to enter the network otherwise, it will not be allowed to enter the network.

In the second step of the above protocol, $m$ is the number of sites in the quantum key or its length, which is provided after a request. For example, consider the $N$ position on the SIM card memory. The network asks for a measurement of positions $n$ to $n + m$. Again, it should be remembered that it is not possible to copy the SIM card due to the quantum no cloning. One of the points that should be considered in working with this protocol and for which we must make a contract is: When the SIM card selects the number of m sites for the key and randomly measures them in one of the $X$ or $Z$ directions. Only the number of $q \in \{0,1,...,m\}$ from the $m$ bases, may be the same for bases at similar positions on the AUC that were previously measured. Now, the question that can be raised is that with which q of the total number of m in a request, the SIM card is identifiable? We know that there are situations such that none of the bases of measurement for AUC and SIMs are the same in a string m. Therefore, in this case, the network should not allow the SIM card to be logged in, or even eliminate it. Before answering this question, we should recall some statistical principles. For a string of $m$ positions that are to be measured in two different directions on those positions, the total number of possible measurements is $2^m$. We also know that the probability that $q$ from the $m$ position has the same measurement bases can be obtained from the following formula.

$$P_q^m = \frac{\binom{m}{q}}{2^m} = \frac{m!}{2^m q!(m-q)!} \qquad (5)$$

With regard to the two selective of the measurement direction, it can observe:

$$\sum_{q=0}^{m} P_q^m = 1 \ \ or \ equivalently \ \ \sum_{q=0}^{m} \binom{m}{q} = 2^m \qquad (6)$$

It can also be seen that this probability for $q_{odd} = \frac{(m \pm 1)}{2}$ and $q_{even} = \frac{m}{2}$ will have its maximum value. This means that the SIM card must, depending on whether m is an Even or Odd number, and as a contract, the number $q_{odd/even}$ of $m$ measures in one direction and the rest in the other direction. Therefore, the network must perform its measurements for each category of m of a total number of $N$ positions with the same contract. So, according to the contract, the network allows the SIM card to be logged only when the number of identical measurement bases is $q = q_{odd/even}$. In this q, the probability of being the same as the bases of measurement is maximal. Of course, if the length of the string m is large and the SIM card is measured randomly, we will naturally get closer to this maximum probability.

**Example:** Suppose that the number of $N = 10^6$ Bell state (or two entangled particles state) as $|\psi\rangle = \frac{1}{\sqrt{2}}(|00\rangle + |11\rangle)$ is shared between the AUC and the SIM card. According to the protocol, the AUC performs measurements on all particles sent from the source, in the arbitrary bases in X or Z direction and records the results classically. Now suppose the request is sent by the SIM card, and the network receives it. The network then announces to the SIM card through the classical channel to measure the particle stored on positions labeled by n = 51 to n + m = 60 and then sends its bases through the classical channel to the network. According to Equation (5), the total number of possible states in the measurement is $2^{10}$. Since $m = 10$ is an even number, the maximum probability for $q_{even} = 5$ occurs and is equal to $P_5^{10} = \frac{252}{1024} = 0.246$. It looks well in identifying a SIM card at once at the request. In accordance with the measurement contract, in order to achieve this lower band in probability, the network and the SIM, for a string of $m = 10$ positions, we need to measure the $q_{even} = 5$ in the direction of X, and the remainder in Z, and it's absolutely random in the sequence X or Z, and independent of each other. Of course, the network already performs its own measurements on the same positions, and stores the results with classical bits along with its corresponding bases in classical memory. In Table 3, the results of measurements of the SIM card and network are presented with the corresponding bases. As it can be seen, wherever the measurement bases are the same (green columns), the results are maintained and used to identify the identity. In this example, the key 1011 is registered. Here we should note that if you try to make a copy of the original SIM card, because of the impossibility of copying the quantum states stored on the SIM card (which have been stored before they are measured), all states, changes and As a result, after measuring the SIM card and comparing it, the results will not be the same. This discrepancy means the existence of a copy of the SIM card, and the network can prevent it from entering.

| | m-State → | $|\psi\rangle_{n+1}$ | $|\psi\rangle_{n+2}$ | $|\psi\rangle_{n+3}$ | $|\psi\rangle_{n+4}$ | $|\psi\rangle_{n+5}$ | $|\psi\rangle_{n+6}$ | $|\psi\rangle_{n+7}$ | $|\psi\rangle_{n+8}$ | $|\psi\rangle_{n+9}$ | $|\psi\rangle_{n+10}$ |
|---|---|---|---|---|---|---|---|---|---|---|---|
| AUC Unit | Measurement Bases | $S_Z$ | $S_X$ | $S_Z$ | $S_Z$ | $S_X$ | $S_Z$ | $S_X$ | $S_Z$ | $S_X$ | $S_X$ |
| | Results | $|0\rangle$ | $|-\rangle$ | $|0\rangle$ | $|1\rangle$ | $|-\rangle$ | $|1\rangle$ | $|-\rangle$ | $|0\rangle$ | $|-\rangle$ | $|+\rangle$ |
| | Classical Bit | 1 | 1 | 0 | 1 | 1 | 1 | 1 | 0 | 1 | 0 |
| MS Unit | Measurement Bases | $S_Z$ | $S_Z$ | $S_Z$ | $S_X$ | $S_Z$ | $S_Z$ | $S_X$ | $S_X$ | $S_Z$ | $S_Z$ |
| | Results | $|1\rangle$ | $|1\rangle$ | $|0\rangle$ | $|-\rangle$ | $|0\rangle$ | $|1\rangle$ | $|-\rangle$ | $|-\rangle$ | $|1\rangle$ | $|0\rangle$ |
| | Classical Bit | 1 | 1 | 0 | 1 | 0 | 1 | 1 | 1 | 1 | 0 |

**Table 3.** Measuring results in the direction of X or Z for the network and SIM card on the entangled Bell state shared between them and recording the results using the classical bits used in the quantum identification protocol using quantum memory.

Figure 2, shows the steps in the quantum authentication protocol of SIM card, using quantum memory as a schematic.

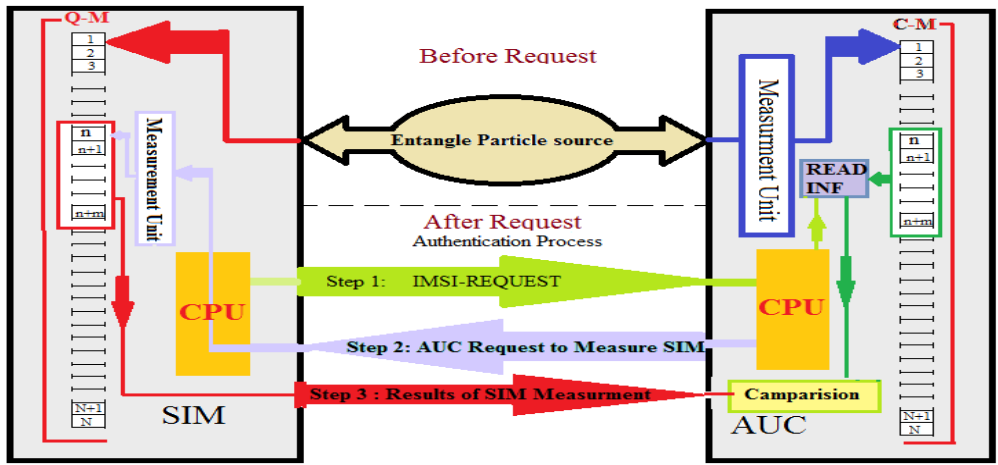

**Figure 2.** A schema of the SIM card identification protocol by the network using quantum memory. In this figure, the CPU stands for the Center of Control Process, the READ INF, the abbreviation for the Read Information which is retrieval center for the classical information stored in classical memory after measured by the network. The C-M stands for Classical Memory and Q-M, an abbreviation for quantum memory quantum memory. The "**Comparison**" is his task to compare the results of measuring the SIM card and memory.

## 5.1. Properties

**(a) Randomness**

Since the protocol uses quantum memory with the ability to store large amounts of information and not be able to be copied from them (no-cloning quantum property), and on the other hand, the AUC collects randomly some measurement results from the SIM cards for identification. Therefore, this kind of identification has a random property.

**(b) Bidirectional authentication**

The authentication in this protocol can be bidirectional. The reason for this can be explained as follows. Although the quantum memory is very large, but limited, so the issue needs a little more curiosity. The problem will begin when a hacker introduces the AUC and attempts to destroy the original SIM card. But it cannot introduce itself to the network instead of the main SIM card. It may only aim to destroy the SIM cards of a mobile company. This is one of the attacks that are happening todays. In this case, the bidirectional authentication can be used. The hacker may do this by sending a subsequent request to retrieve a packet from $K_i$ stored in the original SIM card's quantum memory through a robot. Of course, identifying a hacker from the AUC can be done via bidirectional authentication as usual classic methods available in GSM.

## 5.2. The CNOT Attack

As explained in this section, since in this protocol, the provider of the service to the SIM card, at a distance (just a few meters), initially places entangled particles in the AUC and the SIM card, it seems that here the CNOT attack can not be effective. At the end of this section, It should be noted that, this protocol is based on quantum memory, as it has been stated, before assigning the SIM to the subscriber, a number of entangled particles are shared (in short range) through the quantum channel between the SIM card and the AUC. We know that since this protocol is based on quantum entanglement and it is difficult to send information at long distances, despite the environmental noise, then the protocol attempts to use the classical channel at a later stage (see Figure 2). This can be considered technically and in the language of telecommunication science as the signal loss in long-range airborne transmission due to environmental noise. Therefore, due to the difficulty in implementing the extensive quantum channels of existing telecommunication networks, we propose a preferred protocol so that it can be practically implemented in telecommunications and this protocol focuses on the use of quantum memories which is currently under development. So, in our belief, the use of quantum memory in the SIM card and quantum measurement after the identification request by the AUC, and the transmission of results through the classical channel, seems to be feasible in terms of practical implementation in the existing telecommunication framework.

## 5.3. Noise problem

In practical systems, however, there will always be some back ground noise due to measurement in the detectors and transmission errors. Therefore, in QKD, due to measurements and peripheral noise (which may even arise due to temperature differences in different areas), the part of the key may shift in the transmission of the key. If the error is below a certain threshold, the key can still distill a final secret key using classical protocols for error correction and privacy amplification. If the error is above the threshold, the key is discarded and a new distribution has to be started. On the other hand, as we know, in quantum teleportation, quantum repeaters play an essential role in maintaining the entanglement between particles. A quantum repeater will be an important building block in a future network, since it allows to interconnect different network nodes. In a quantum repeater, two particles of independent entangled pairs are combined within a Bell state measurement (BSM), such that the entanglement is relayed onto the remaining two particles. This process is called entanglement swapping and will eventually allow to overcome any distance limitations in a global-scale network. However, in order to efficiently execute entanglement swapping, it has to be supplemented with an entanglement purification step requiring quantum memories. In our paper, since the both protocols are preceded by secure sharing of the entangled state. Therefore, there may be concerns about the proper transmission of entangled particles. But it can be said, given the protocols presented here, it can be seen that in the second protocol based on quantum memory, there is no problem with the transmission of entangled states, at close intervals provided by the mobile service provider [21], [24-25]. This can be accomplished by sending two entangled photons and fixing their states in solid quantum memory, which is constructed using the theory of Dots. In the case of the first protocol, based on the transmission of three entangled photons at distances through the air channel, as we know, free-space quantum communication, is one of the problems in QKD. In past years, several experiments have been performed over various distance [30-37]. Also, it is important to overcome the noise problem in transmitting information.

## 6. Summery

In this paper, we tried to propose quantum authentication protocols for mobile networks using quantum cryptography. With use these protocols, it is possible to authenticate copied SIM cards that are sometimes provided to eavesdrop on the original SIM cards and prevent them from entering the network. The first protocols were presented using the source of three entangled particles with GHZ-state based on the E91 protocol. It was observed that whenever the original SIM card and its copy simultaneously request entry to the network then the network can detect the existence of a copy and then can delete both of them. The second protocol presented is based on the use of quantum memory embedded on the SIM card, and used of a source of two particles entangle state with Bell-state. In our opinion, these two protocols can be a good way to authenticate the SIM cards that been copied, in the next generation of mobile networks.

**ACKNOWLEDGMENTS**

Acknowledge support from their respective budget, Azad Islamic university, Ahvaz Branch, Ahvaz. We thank the anonymous referee whose suggestions have contributed toward the improvement of this report.